# Astro2020 Science White Paper

# Ultraviolet Spectropolarimetry as a Tool for Understanding the Diversity of Exoplanetary Atmospheres

**Thematic Areas**: ☒ Planetary Systems   ☐ Star and Planet Formation
☐Formation and Evolution of Compact Objects ☐Cosmology and Fundamental Physics
☐Stars and Stellar Evolution  ☐Resolved Stellar Populations and their Environments
☐Galaxy Evolution        ☐Multi-Messenger Astronomy and Astrophysics

**Principal Author:**
Name:       Luca Fossati
Institution:    Space Research Institute, Austrian Academy of Sciences
Email:      luca.fossati@oeaw.ac.at
Phone:      +43 316 4120 601

**Co-authors:** (names and institutions)
L. Rossi (LATMOS/CNRS, FR), D. Stam (TU Delft, NL), A. García Muñoz (TU Berlin, DE), J. Berzosa-Molina (TU Delft, NL), P. Marcos-Arenal (UCM, ES), J. Caballero (CSIC-INTA, ES), J. Cabrera (DLR Berlin, DE), A. Chiavassa (Nice Observatory, FR), J.-M. Desert (UVA, NL), M. Godolt (TU Berlin, DE), L. Grenfell (DLR Berlin, DE), C. Haswell (Open Univ., UK), P. Kabath (ASU, CZ), K. Kislyakova (Uni. Vienna, AT), A. Lanza (INAF-CT, IT), A. Lecavelier (IAP, FR), M. Lendl (IWF, AT), E. Pallé (IAC, ES), H. Rauer (DLR/TU/FU Berlin, DE), S. Rugheimer (Univ. Oxford, UK), A. Vidotto (TCD, IE)

**Abstract**:
The polarization state of starlight reflected by a planetary atmosphere uniquely reveals coverage, particle size, and composition of aerosols as well as changing cloud patterns. It is not possible to obtain a comparable level of detailed from flux-only observations. Furthermore, polarization observations can probe the atmosphere of planets independently of the orbital geometry (i.e., transiting and non-transiting planets). We show that a high-resolution spectropolarimeter with a broad wavelength coverage, particularly if attached to a large space telescope, would enable simultaneous study of the polarimetric exoplanet properties of the continuum and to look for and characterize the polarimetric signal due to scattering from single molecules.

# 1. Introduction

The discovery of exoplanets opened an exciting new field in astrophysics and planetary sciences. Almost 4000 exoplanets are currently known and thousands more are expected to be found in the near future. As a natural consequence of this success, the research focus is shifting from planet detection to atmospheric characterization, as shown, e.g., by the fact that TESS and PLATO focus on searching transiting planets orbiting nearby, bright stars, specifically to enable atmospheric characterization.

Despite the still limited sample and data quality, observations clearly indicate that exoplanetary atmospheres are extremely diverse, ranging from planets showing strong atomic and molecular features to others which are nearly featureless, indicating the possible presence of aerosols (e.g., Sing et al. 2016). These data are mostly obtained through transmission spectroscopy, which however provides no or very limited information on the microphysical properties (particle composition, size, shape, number density) of clouds and haze layers. Broad wavelength coverage, high-resolution spectropolarimetry of starlight that is reflected by an exoplanet promises to be a game changer in our quest towards better understanding the atmospheres of these alien worlds, particularly if carried out simultaneously with total flux measurements.

# 2. Spectropolarimetry of exoplanetary atmospheres
## 2.1 General considerations

Polarimetry is a powerful tool for studying (exo)planets. Since reflections in general polarize light, the starlight reflected by a planet is polarized whereas, integrated over the stellar disk, the light of most main-sequence stars is unpolarized (Kemp et al. 1987; Cotton et al. 2016, 2017). The degree of linear polarization (P; the polarized-to-total flux ratio) is sensitive to the properties of the scattering particles, the atmospheric structure, and, if present, reflection by the planetary surface. It is however relatively insensitive to a number of instrumental effects, and can be determined with high accuracy ($\approx 10^{-5}$-$10^{-6}$). Several instruments looking for long-period planets through direct imaging (e.g., SPHERE, GPI) employ polarization to increase the star-planet contrast and facilitate planet detection. Polarimetry is considered for good reasons to be a primary tools for detecting water and oxygen on directly imaged exoplanets (e.g., Bailey 2007; Karalidi et al. 2012; Miles-Páez et al. 2014; Fauchez et al. 2017; Rossi & Stam 2017).

The line and continuum polarization state of starlight that is reflected by a planet depends on the star-planet-observer phase angle (i.e., the angle between the star and the observer as seen from the center of the planet) and is sensitive to the optical properties of the planetary atmosphere and surface (Fig. 1; e.g., Seager et al. 2000; Stam et al. 2003, 2004, 2006; Rossi & Stam 2018). For example, Hansen & Hovenier (1974) and Tomasko & Smith (1982) used disk-integrated polarimetry of Venus and Titan, respectively, to derive the composition and size of the cloud particles. Due to the lack of sensitivity, total flux measurements cannot provide this information.

The degree and direction of polarization of the starlight scattered and reflected by a planet depends on the illumination, viewing geometry, optical properties of the atmospheric constituents, reflection properties of the planetary surface, and wavelength. Several processes can polarize light in planetary atmospheres: Rayleigh diffusion, hazes, aerosols, etc. The Rayleigh scattering cross section varies roughly as $1/\lambda^4$, making P larger at shorter wavelengths, while scattering by larger aerosols varies in a more complex way, and the transition between the two processes depends on the aerosol's microphysical properties and atmospheric pressure. The polarization phase function of a planet carries the characteristic signatures of the light that is singly scattered by gas, aerosol, and/or cloud, and these, unlike for the total flux, are not erased by multiple scattering. Therefore, polarization observations are compelling for advancing our understanding of extrasolar planetary atmospheres.

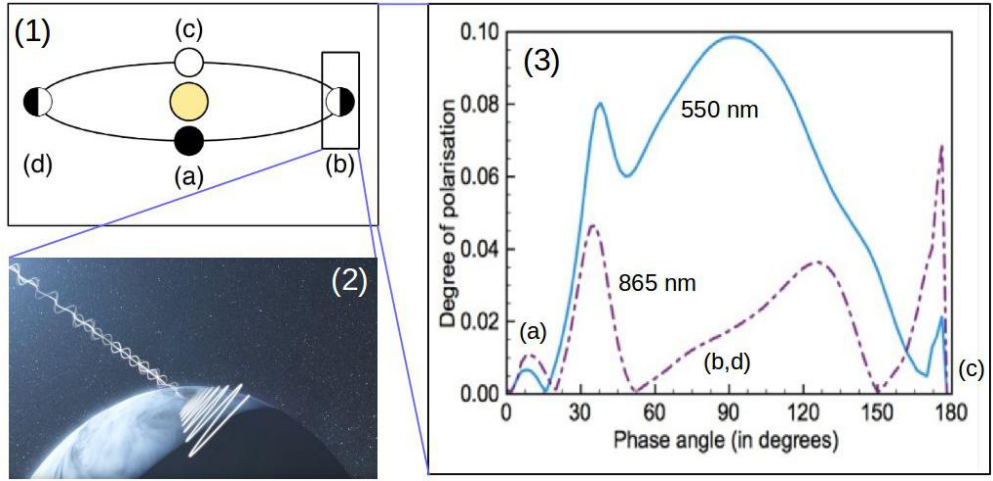

*Figure 1. Panels (1) & (2): schematic of a planetary system in which the unpolarized starlight becomes polarized through reflection by a planetary atmosphere. Panel (3): degree of polarization (P; in %) of a cloudy Earth-like planet as a function of planetary orbital phase, labeled as in panel (1). The different lines are for different wavelengths.*

**2.2 Dependence on orbital configuration and planetary parameters**
The P of a celestial body can reach several tens of percent (e.g., Titan), in particular when the planetary phase angle is close to 90° (i.e., quadrature; Seager et al. 2000, Saar & Seager, 2003, Hough & Lucas 2003; Stam 2008). However, for spatially unresolved planets, P will depend strongly on the stellar background signal, which adds unpolarized light to the planetary signal hampering polarization measurements. At phase angles close to 0° or 180°, P is close to zero (Fig. 1). Thus, polarimetry is ideally suited to characterize the atmospheres of non-transiting planets.

Figure 2 shows the strong dependence of P upon surface albedo and cloud coverage, which play a key role in the energy budget of a planetary atmosphere and in turn influence the atmospheric composition (e.g., Karalidi et al. 2011; Karalidi & Stam 2012; Miles-Páez et al. 2014). Furthermore, detecting irregular temporal variations in polarization could indicate changing cloud patterns (i.e., weather), constraining also heat transportation and distribution (e.g., García Muñoz & Mills 2015). Polarization is sensitive to the microphysical properties of planetary atmospheres. Polarization allows one to derive size and shape of the particles forming the clouds; for example, the single scattering properties of ice (crystal) particles are usually very different from those of liquid (spherical) particles (e.g., Goloub & Arino 2000). This is critical information for understanding how clouds develop and evolve. The sensitivity of polarization to the microphysical properties of the scatterers in a planetary atmosphere make this a key tool for breaking degeneracies which can arise for flux-only observations.

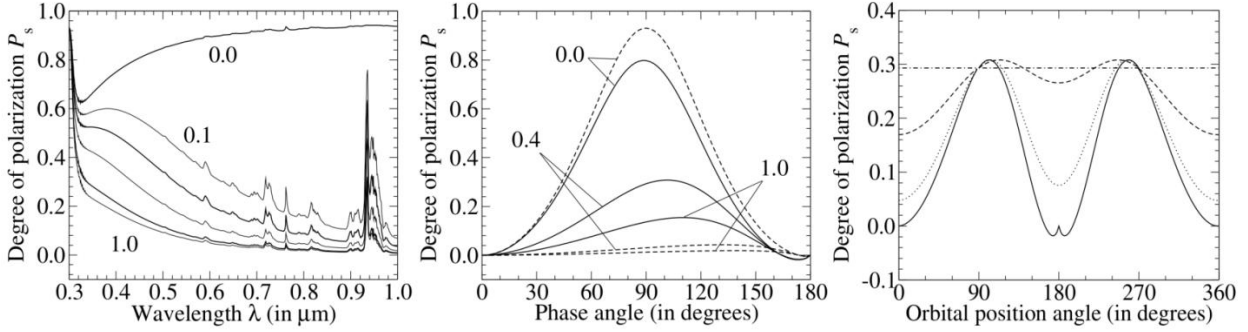

*Figure 2.* Left: degree of polarization (P) of starlight reflected by model planets with Earth-like clear atmospheres vs wavelength and surface albedo (0.0 to 1.0). Middle: P vs phase angle for surface albedos of 0.0, 0.4, and 1.0 and at wavelengths of 0.44 (solid) and 0.87 µm (dashed). Right: P vs orbital position angle at a wavelength of 0.44 µm and for a surface albedo of 0.4, for orbital inclination angles of 0° (dot-dashed line), 30° (dashed line), 60° (dotted line), and 90° (solid line). From Stam (2008).

Since polarization simulations of starlight reflected by an exoplanet extending to the ultraviolet (UV), where P is believed to particularly large, have never been performed before, we employed the PYMIEDAP code (Rossi et al. 2018) to show the exquisite sensitivity of UV polarization measurements to the physical properties of planetary atmospheres. Figure 3 shows flux (top row) and P (bottom row) as a function of phase angle at three UV wavelengths (150, 200, and 300 nm) for a HD189733b-like exoplanet hosting a clear atmosphere (left) and an atmosphere dominated by $NH_3$ (middle) or $MgSiO_3$ (enstatite; right) aerosols. The three curves have very distinct behaviors, particularly as a function of wavelength, clearly indicating the potential of UV spectropolarimetry for gaining critical information on planetary atmospheres, complementary to that provided by flux-only measurements.

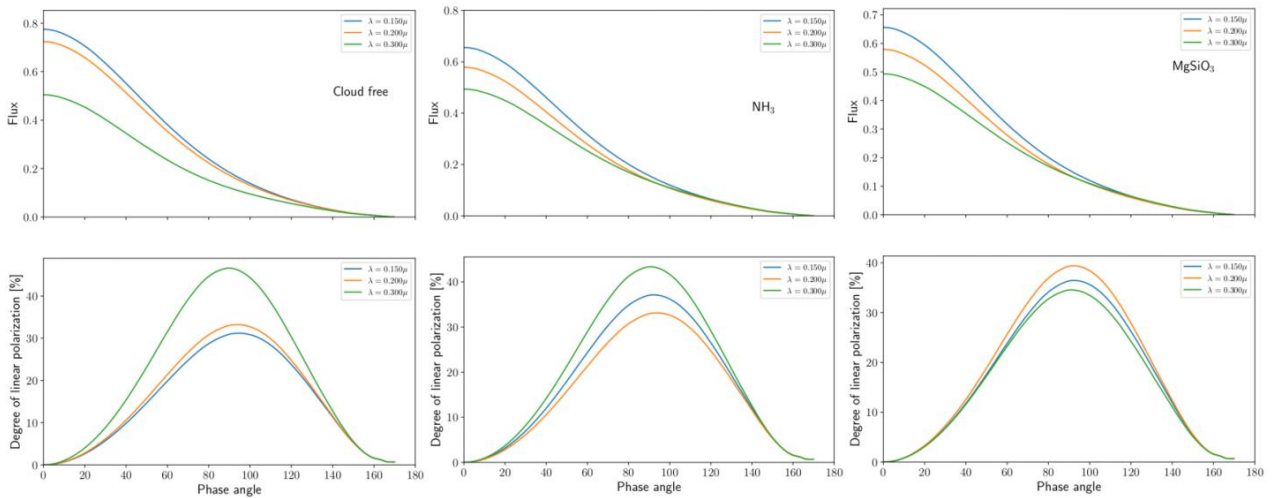

*Figure 3.* Flux (top) and degree of polarization (P; bottom) as a function of phase angle at three UV wavelengths for a HD189733-like exoplanet hosting a clear atmosphere (left) and an atmosphere dominated by $NH_3$ (middle) or $MgSiO_3$ (right) aerosols.

Previous efforts have concentrated on low spectral resolution (<100) instruments trying to drive up the signal-to-noise ratio. A major difficulty of broadband polarimetry is the need to correct for spurious signals introduced by the interstellar medium, the Earth's atmosphere, or the telescope. It has been recently shown that high resolution (>100,000) spectropolarimetric observations are possible and are correspondingly powerful in constraining atmospheric properties (García Muñoz 2018). The polarization of a spatially-unresolved exoplanet can be detected by cross-correlating high-dispersion linear polarization and intensity spectra of the star-planet system. This technique allows one to separate polarization contributions originated at different radial velocities that might be proved to be useful for disentangling the planet's polarization signal from other contributions (e.g., interstellar medium). It may also prove useful to remove the effect of systematics provided that their signal is either a smooth function of wavelength or mimics the stellar spectrum without shifts in wavelength. In other words, dispersing the polarized light provides a natural way of separating the planet's signal (which is Doppler shifted) from other polarization signals introduced in the photon path.

Cross-correlating high-resolution polarimetric spectra requires knowledge of the opacities of the molecules of interest at high resolution. This information is already available for several species relevant for exoplanet studies, particularly at optical and infrared wavelengths, but many are still lacking in the ultraviolet (e.g., $O_2$, $O_3$), where the polarization signal is typically strongest. However, work in this direction is being carried out and the next few years will see major improvements in this respect.

## 3. Instrument requirements

With the currently available technology and what is foreseeable in the near future, spectropolarimetric observations to characterize the close-in, unresolved planets would be realistically possible only for planets larger than Neptune. Smaller planets would require a prohibitive number of observations to achieve an acceptable signal-to-noise ratio, demanding unrealistic stability requirements over long timescales.

To make best use of the available data analysis techniques, and thus to enable measurements at both high and low resolution, the instrument require a resolution greater than 100,000. García Muñoz (2018) shows that lower spectral resolutions disentangle the planetary and non-planetary polarization signals less efficiently. A high resolving power is therefore critical to take advantage of cross-correlation techniques, still allowing measurements at low resolution. An efficiency of at least 50% and a high stability would be also desirable. Physical considerations (e.g., Rayleigh scattering) suggest that the near-UV (170-350 nm) and the blue part of the optical waveband are valuable regions to cover for polarimetric observations. Therefore, the ideal instrument would cover these regions, possibly extending to the optical and near-infrared to exploit the cross-correlation technique on the wealth of molecules present there. The need to cover the near-UV wavelength band and the required high stability, given also by the high spectral resolution, imply that the instrument would be flown in space. Furthermore, the space-based platform gives freedom on the scheduling of the observations, which is an important point for exoplanet observations that need to be conducted along the planetary orbital phase and possibly close in time to minimize systematic uncertainties.

To identify the aperture size of the telescope that is required to characterize the atmospheres of short-period (≈2 days) exoplanets, we computed the signal-to-noise ratio in the 250-400 nm spectral region as a function of telescope aperture obtained over 3 hours of shutter time and with 5 nm binning. We employed a telescope-spectropolarimeter system providing an average efficiency of 50%, which is reasonable considering the available technology. Figure 4 presents the results considering G2V, F5V, and A0V host stars.

By focusing on the near-UV, Fig. 4 shows that detecting and characterizing the atmospheres of close-in giant planets orbiting G-type stars would require a telescope with an aperture greater than 9-10 m. Because of the increasing fluxes of the host stars, measuring polarization becomes easier with increasing the temperature of the host stars. Figure 4 clearly shows that close-in gas giants orbiting A-type stars would be excellent targets. Furthermore, a telescope aperture larger than 10 m would enable to detect the polarized reflected light even for systems as far as 500 pc.

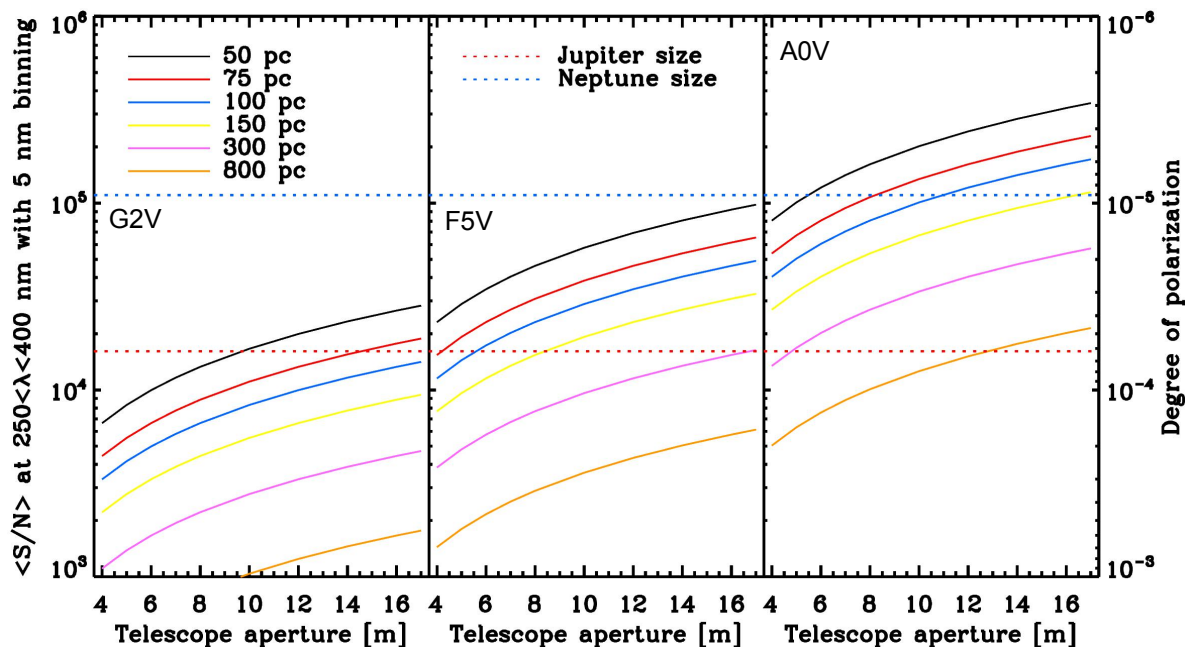

*Figure 4.* *Average signal-to-noise ratio of the data (i.e., inverse of the measurable polarization level) between 250 and 400 nm obtained with 3 hours of shutter time and 5 nm binning as a function of the distance to the target star (line colour) and size of the telescope primary mirror. Results are for a G2V (left), F5V (center), and A0V (right) star. The red and blue dotted horizontal lines indicate the maximum degree of polarization (P) calculated respectively for a Jupiter-radius and Neptune-radius planet in a 2-day orbit around the host star. We considered that each planet gives a maximum P of 30% and that the efficiency of the telescope-instrument system is on average 50%.*

Thanks to the unique possibility given by polarimetry to characterize the atmospheres also of non-transiting exoplanets and by the large number of nearby systems discovered so far by both radial velocity and transit surveys, several targets would be already available for in-depth characterization. As of today, a target list composed of planets with masses larger than Saturn (to ensure Jupiter-like planetary radii; e.g., Hatzes & Rauer 2015), orbital periods shorter than 3 days, and orbiting stars hotter than the Sun and closer than 200 pc would comprise about 20 systems. Furthermore, a number of additional (possibly better) targets can be expected to become available soon following the results of the Gaia, TESS, and PLATO missions and radial velocity surveys.